\documentclass{ledger}
\newcommand{\thefirstpagenum}[0]{X}
\newcommand{\thelastpagenum}[0]{X}
\newcommand{\theyear}[0]{20XX}
\newcommand{\thevol}[0]{X}
\newcommand{\thedoi}[0]{DOI 10.5195/LEDGER.\theyear.XXX}
\newcommand{\ledgerpages}[0]{\thefirstpagenum-\thelastpagenum}

\title{Towards a zk-SNARK compiler for\\Wolfram language}
\author{
  Armando B. Cruz
  \thanks{A. B. Cruz (aleph\_g@ciencias.unam.mx) Department of Mathematics, Faculty of Science, Universidad Nacional Autónoma de México, Mexico City, Mexico.}
}
\pretitle{
  \centering \selectfont RESEARCH ARTICLE \par 
  \fontsize{24pt}{28pt}\selectfont
} % Title is centered and at 24pt

\begin{document}

\maketitle

\thispagestyle{pagefirst}

\begin{abstract}
  Zero-knowledge proofs (zk-Proofs) are communication protocols by which a prover can demonstrate to a verifier that it possesses a solution to a given public problem without revealing the content of the solution. Arbitrary computations can be transformed into an interactive zk-Proof so anyone is convinced that it was executed correctly without knowing what was executed on, having huge implications for digital currency. Despite this, interactive proofs are not suited for blockchain applications but novel protocols such as zk-SNARKs have made zero-knowledge ledgers like Zcash possible.
  This project builds upon Wolfram's ZeroKnowledgeProofs paclet and implements a zk-SNARK compiler based on Pinocchio protocol.

  %AUTHOR: keywords are OK to show for Review article, will be hidden and added to metadata for publication
\end{abstract}

\section{Introduction} {
  In this work, the author builds upon Wolfram's ZeroKnowledgeProofs paclet to implement, for the first time, the first four stages of a code-to-zk-SNARK compiler based on Pinocchio protocol\cite{Parno2013} for programs written in a subset of Wolfram language.

  Zero-knowledge proofs (zk-Proofs) are communication protocols by which a prover can convince a verifier that a computation was executed correctly without the verifier having to execute the computation itself and without knowledge of the inputs. zk-SNARKs are a type of non-interactive zk-Proof that allows scalable simultaneous verifications and that are well suited for blockchain ledgers based on \textit{proof-of-work}, see Section~\ref{sectionZKBlockchain}. This have some important applications such as Completely anonymous ledgers Subsection~\ref*{subsectionZcash}; Fraudless democratic elections Subsection~\ref*{subsectionDemocracy}; Zero-knowledge Wolfram code review Subsection~\ref*{subsectionZKPeers}.
  
  \subsection{Interactive zk-Proofs paclet} {
    Interactive zero-knowledge proofs defined in Wolfram language, are communication protocols by which a prover $P$ that has a \textit{private solution} to a \textit{public problem} of complexity class NP can convince a verifier $V$ that it possesses such solution without revealing its content \textit{i.e.} the verifier gains zero-knowledge.
    \\
    In the common non-zero-knowledge case, $P$ will simply share the solution with $V$ and then $V$ will verify this by running the polynomial time verification algorithm. In order to achieve zero-knowledge, $P$ must transform the public problem into several equivalent problems for which $V$ will send repeated challenges to $P$ that can be solved efficiently only if $P$ posses said solution. This is implemented in the following protocol:
    \begin{enumerate}
    \item{} $P$ generates an isomorphism that ciphers the public problem into an equivalent NP problem called the \textit{public cipher problem}, the isomorphism must meet:
      \begin{itemize}
        \item Is a one-way function, that is, impossible to revert.
        \item The private solution to the original problem is also ciphered into a solution to the new public cipher problem.
      \end{itemize}
      Both the isomorphism and the cipher solution are shared with $V$ as witness of the solution.
    \item{}
      To verify the witness, $V$ must request $P$ to show either the isomorphism used to cipher, or the solution to the equivalent public cipher problem. $V$ can only ask for one of those since having both is all it needs to recreate the original private solution making this protocol non-zero-knowledge.
      $V$ cannot deduce the private solution from partial information, because the isomorphism is generated independently and it's impossible to revert the cipher solution.
    \item{}
      If $P$ returns an incorrect response, $V$ can be sure that the claimed solution is forged and rejects the proof. On the other hand, if the response is valid, then either $P$ has the solution, or it was lucky and $V$ asked for a forged response, the latter case has a 50\% probability of success, and decreases exponentially with the number of rounds this interaction is repeated.
    \end{enumerate}
    Wolfram's ZeroKnowledgeProofs paclet\cite{cruz2022} implements this protocol using two distinct types of NP-complete public problems shown in Table~\ref{TableInteractiveProtocols}.
    
    \begin{table}
      \begin{center} 
      \scalebox{0.6}{ 
        \begin{tabular}{p{5cm}p{18cm}}%{l|}%{|>m{5cm}|>m{15cm}|}
        %\rowcolor{orange}
          \hline
            \textbf{Protocol}  & \textbf{Description} \\
          \hline
            HamiltonianCycle & The public problem consists in finding a hamiltonian cycle in a public graph. The problem is ciphered by graph isomorphisms.\\
          \hline
            SAT & The public problem consists in finding a boolean vector that satisfies a public boolean function expressed in SAT-3 format. The problem is ciphered by a permutation of the boolean blocks and their signs.\\
          \hline
        \end{tabular}
      }
      \caption{Implemented NP complete public problems in interactive protocol} \label{TableInteractiveProtocols}
      \end{center}
    \end{table}
  }

  \subsection{zk-Proofs for arbitrary computations}\label{subsectionArbitraryProof} {
    The complexity class NP-complete is composed of problems for which any other problem of class NP can be transformed into an instance of this NP-complete problem. Cook-Levin Theorem\cite{Gamboa2004} states that the boolean satisfiability problem (SAT) is of class NP-complete.
    This implies that the above protocol can be extended to verify that an arbitrary computation was executed correctly by $P$ without $V$ executing it nor knowing what input was executed, all it's needed is a way to \textit{compile} the program and its solution into an equivalent SAT problem and generate its interactive zero-knowledge proof. One way to achieve this is to write the boolean circuit for the computation using only OR, AND \& NOT gates and encode it into an equivalent boolean function.
  }
}
  
\section{Zero-knowledge blockchain}\label{sectionZKBlockchain} {
  \subsection{Limitations of interactive zk-Proofs} {
    The fact that any program can be verified with an interactive zero-knowledge proof has several implications for digital currency, for example in a centralized digital currency scheme, transactions between accounts could be verified by the central authority without it knowing the amounts transferred between them. The main disadvantages of interactive proofs are:
    \begin{itemize}
      \item The witness size and verification time grows linearly with the number of rounds.
      \item The cipher computation time grows quadratically with the number of rounds and the number of verifiers.
      \item In every verification $P$ must be an active part of the communication, making nonscalable for multi-verification systems.
    \end{itemize}
    In essence, interactive proofs are not suited for blockchain ledgers based on \textit{proof-of-work} where several miners must verify transactions simultaneously before finding the next block solution.
    \\
    Advancements in zero-knowledge\cite{Groth2010} developed a novel method called zk-SNARK (zero-knowledge succinct non-interactive arguments of knowledge) a non-interactive zk-Proof protocol having the \textit{public verifier} property, that is that anyone can verify the proof without interacting with the prover anew.
    The principle that makes zk-SNARKs possible is the same as the one described in Subsection~\ref{subsectionArbitraryProof}: An arbitrary computation must be compiled into a \textit{Quadratic Arithmetic Program} (QAP) an NP-complete problem introduced by  Gennaro, Gentry, Parno and Raykova\cite{Gennaro2013}, and then based on homomorphic hiding techniques a zk-SNARK proof is generated.
  }

  \subsection{Zcash algorithm in a nutshell}\label{subsectionZcash} {
    Zcash algorithm\cite{Sasson2014} introduced the application of zk-SNARKs to anonymous payments in the form of \textit{shielded transactions} that can be fully encrypted on the Zcash blockchain, yet still be verified by miners under the network's consensus rules.
    \\
    At high level this is done by compiling each transaction along with the network's rules into a QAP and its corresponding zk-SNARK proof whose validity holds if and only if the transaction verifications were done correctly (\textit{p.e} the input values sum to the output values, the sender has the authority to spend the input amount, etc.) and revealing no information to the miners about the input values it performed the calculations on (\textit{p.e} wallet addresses, amount transferred, etc.). A more indepth explanation of the Zcash protocol can be found in this research paper\cite{Banerjee2020}.
  }

  \subsection{Decentralized democratic elections}\label{subsectionDemocracy} {
    Zero-knowledge blockchains like Zcash could be particularly useful for noncurrency applications where sensitive information must remain private yet verifiable by a community, like decentralized democratic elections; here each democratic party has a publicly known wallet address and each voter a unique private wallet. Elections would be carried as follows:
    \begin{itemize}
      \item At the start of the elections each voter's wallet is created and given one unit of currency.
      \item A vote is defined as a shielded transaction of one unit from a voter wallet to a party wallet, transactions amounts can only be of one unit, this ensures at most one vote per wallet.
      \item All voters must participate in the zero-knowledge blockchain mining competition.
      \item Party wallets balance must remain private until the end of elections.
    \end{itemize}
    All this constraints are to be compiled as zk-SNARKs and added to the network's rules and mining algorithms. By the end of the elections the party wallet with the most currency (votes) corresponds to the winning party. This protocol ensures the privacy of the voter's choice and the collective consensus of a fraudless election. 
  }

  \subsection{Zero-knowledge Wolfram code review}\label{subsectionZKPeers} {
    Another possible noncurrency application of zk-SNARKs to scientific research is the possibility of proving to a reviewer the correctness of research results obtained in Wolfram language without sharing the author's code, models or parameters. 
    This could be particularly useful for machine learning research where assertions about the performance of a trained model are made, but verifying this usually involves sharing the model weights.
  }

  \begin{figure}
    \centering\captionsetup{justification=centering}
    \includegraphics[width=0.85\textwidth]{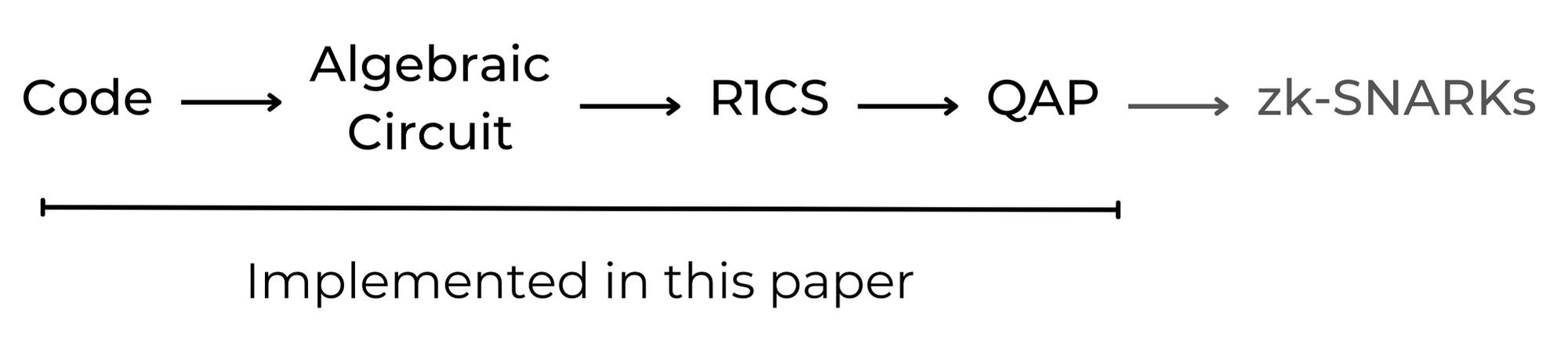}
    \caption{Pinnochio compilation pipeline.}
    \label{figPinocchioPipeline}
  \end{figure}
}

\section{Pinnochio protocol implementation} {
  Several zk-SNARK compilers have been developed for a variety of languages, 
  \textit{libsnark} for C++, 
  \textit{gnark} for Go, 
  \textit{halo2} for Rust this is the library utilized by Zcash zero-knowledge proving system in the sixth major update in May 2022.
  The Pinocchio protocol\cite{Parno2013} introduced by Parno and Gentry proposed an efficient end-to-end toolchain for compiling a program written in a subset of C into a QAP that can be easily transformed into a zk-Proof with efficient verification time. Conversion from code to proof is done in five steps as shown in Figure~\ref{figPinocchioPipeline}. 
  \\
  In this paper, the author implemented the first four stages of this pipeline for programs written in a subset of Wolfram language and presents an explanation of each step along with an example taken from the paclet code. The full code can be found in Appendix~A.

  \subsection{Program definition} {
    Suppose $P$ wants to convince $V$ that it knows a 3-coloring for the graph $G=(U,E)$ where $U=\{u_1,u_2,u_3,u_4,u_5\}$ and
    \begin{equation}
      E=\{
        (u_1,u_2),(u_1,u_3),(u_1,u_4),(u_1,u_5),
        (u_2,u_5),(u_2,u_3),(u_3,u_4),(u_4,u_5)
      \}
    \end{equation}
    To generate a QAP $P$ must come up with a set of polynomials and a corresponding output condition that encapsulates the verification algorithm of a 3-coloring on this graph. For example the program 
    $(f_1(c_1,\ldots,c_5) \neq 0,
    f_2(c_1,\ldots,c_5) = 0)$ where
    \begin{equation}\label{eqF1}
      f_1(c_1,\ldots,c_5)=
        (c_1-c_2)(c_1-c_3)(c_1-c_4)(c_1-c_5)
        (c_2-c_5)(c_2-c_3)(c_3-c_4)(c_4-c_5)
    \end{equation}
    \begin{equation}\label{eqF2}
      f_2(c_1,\ldots,c_5)=
        \textstyle{\sum_{i=1}^{5}}
        (1-c_i)(2-c_i)(3-c_i)
    \end{equation}
    the first condition ensures no two adjacent vertexes have the same value and the second one ensures there are at most three colors. 
  }

  \subsection{Algebraic circuit} {
    Once the program has been defined, an arithmetic circuit is constructed for each polynomial. Arithmetic circuits are analogous to boolean circuits but instead of boolean gates (AND, OR, NOT), arithmetic gates (Plus, Times) are used.
    \begin{figure}[t!]
      \centering\captionsetup{justification=centering}
      \includegraphics[width=\textwidth]{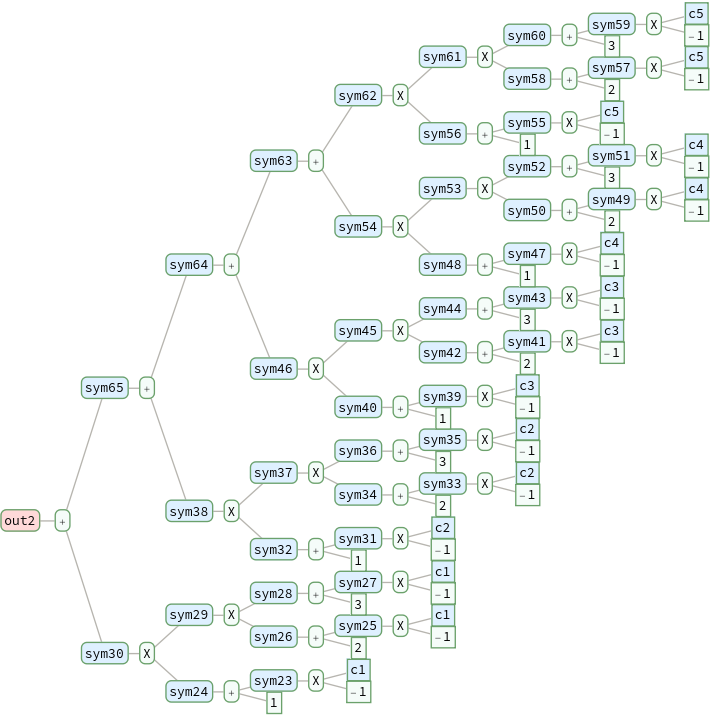}
      \caption{Arithmetic circuit of $f_1$.}
      \label{figArithmeticCircuit}
    \end{figure}
    \\
    Figure~\ref{figArithmeticCircuit} presents the circuit of the first polynomial, in this diagram blue cells correspond to inputs and internal computations, green cells to constant values and gates, and the red cell is the output in this case $out_1$ must be different than 0. An association $t$ is said to be a solution to the circuit if it assigns to each blue cell the numerical value resulting from the gate computation and if the output cell satisfies the program condition.
  }

  \subsection{Quadratic arithmetic program} {
    Both the circuits and their solution have all the information needed to verify that $P$ has a valid three coloring of the graph, but still needs a way to cipher them into a zk-Proof, the next step will generate for each input and each internal symbol $i\in I$ three special polynomials $(v_i,w_i,k_i)$ that together encode the complete circuit. To cover for constant values in the circuit, the internal symbol \textit{one} is introduced and its triplet is calculated by the same rules. 
    \\
    The degree of these polynomials is at most the combined number $N$ of arithmetic gates in the circuits, for this particular example $N=66$, most of them will reach the upper bound making computation for $P$ extensive, but this is a worthy tradeoff, in the end $P$ will be able to generate a zk-Proof that is easily verifiable any number of times without its interaction.
    \\
    The properties that the polynomial triplets must satisfy for all $1\leq d\leq N$ are:
    \begin{itemize}
      \item{} $k_i(d)=1$ if and only if symbol $i$ is the output to the $d^\text{th}$ arithmetic gate.
      \item{} $v_i(d)\neq0$ or $w_i(d)\neq0$ if and only if symbol $i$ is an input of the $d^\text{th}$ arithmetic gate.
      \item{} $t$ is a solution to the arithmetic circuits if and only if 
      \begin{equation}\label{eqT}
        F(d)=v(d)w(d)-k(d)=\sum t(i)v_i(d)\sum t(i)w_i(d)-\sum t(i)k_i(d) = 0
      \end{equation}
    \end{itemize}
    Notice that Equation~\ref{eqT} relates all triplets of polynomials and the circuit solution into a unified polynomial $F$. When $t$ is a valid circuit solution $R$ has at least $N$ roots, and it can be factored by $T(x)=\prod(x-d)$ like $F(x)=H(x)*T(x)$, $T$ is called the \textit{target polynomial}.
    The Quadratic arithmetic program generated by this arithmetic circuits is defined as the tuple 
    $Q=(\{v_i\},\{w_i\},\{k_i\},T)$.
    It's important to mention that all polynomials, inputs and intermediate symbol calculations are made over a finite field $\mathbb{F}$, and will be referring to its multiplicative group as $\mathbb{G}$.
  }

  \subsection{Verify computations with QAP} {
    Suppose $P$ is trying to deceive $V$ into accepting an incorrect solution $t^\prime$ to a publicly known algebraic program as valid. If the problem was previously compiled into a QAP, $Q=(\mathcal{V},\mathcal{W},\mathcal{K},T)$ and $V$ agrees on the correctness of the compilation, then by the characteristic property of QAP, $F(x)=v(x)w(x)-k(x)$ will not be divisible by $T(x)$ when substituting $t^\prime$ \textit{i.e.} that for any polynomial $H$, $F\neq HT$ and so $F(x)=H(x)T(x)$ on at most $2N$ points. Thus, if the order of the field is much larger than $2N$, the probability that the equality holds on a randomly chosen $s\in\mathbb{F}$ is very small.
    \\
    Let $g$ be a generator of the multiplicative group $\mathbb{G}$ and $E:\mathbb{G}\times \mathbb{G}\rightarrow \mathbb{G}$ be a non-trivial bilinear map. Suppose $P$ wants to let $V$ know that it possesses the solution to an algebraic program.
    \begin{enumerate}
      \item{} $V$ will randomly generate 
      $r_v, r_w, s, \alpha_v,\alpha_w,\alpha_k, \beta,\gamma \in\mathbb{F}$
      and let $r_k=r_v\cdot r_w, g_v=g^{r_v}, g_v=g^{r_v}, g_k=g^{r_v}$, and construct the public \textit{evaluation key}
      \begin{eqnarray*}
        (
          \{g^{s^d}\}_{d\leq N},
          \{g_v^{v_i(s)}\}_{i\in I},
          \{g_w^{w_i(s)}\}_{i\in I},
          \{g_k^{k_i(s)}\}_{i\in I},
          \{g_v^{\alpha_v v_i(s)}\}_{i\in I},
          \\
          \{g_w^{\alpha_w w_i(s)}\}_{i\in I},
          \{g_k^{\alpha_k k_i(s)}\}_{i\in I},
          \{g_v^{\beta v_i(s)}g_w^{\beta w_i(s)}g_k^{\beta k_i(s)}\}_{i\in I}
        )
      \end{eqnarray*}
      and the public \textit{verification key}
      \begin{eqnarray*}
        (
          g,g^{\alpha_v},g^{\alpha_w}g^{\alpha_k},
          g^{\gamma},g^{\beta\cdot\gamma},g_k^{T(s)},
          \{g_v^{v_i(s)},g_w^{w_i(s)},g_k^{k_i(s)}\}_{i\in I},
        )
      \end{eqnarray*}
      Both keys are published into a public ledger shared with $P$, it's impossible to retrieve the randomly generated values from the keys due to the discrete logarithm problem. 
      \item{} $P$ gets the evaluation key and using this values computes the witness key and publish it into the same public ledger
      \begin{eqnarray*}
        (
          g_v^{v(s)},g_w^{w(s)},g_k^{k(s)},g^{H(s)}
          g_v^{\alpha_v v(s)},
          g_w^{\alpha_w w(s)},
          g_k^{\alpha_k k(s)},
          g^Z=g_v^{\beta v(s)}g_w^{\beta w(s)}g_k^{\beta k(s)}
        )
      \end{eqnarray*}
      \item{} Any other verifier $V^\prime$ with access to the verification and witness keys can verify the computation in three steps
      \begin{itemize}
        \item{} To assert that T divides F, checks
        $
        E(g_v^{v(s)},g_w^{w(s)}) = 
        E(g_k^{T(s)},g^{H(s)})E(g_k^{k(s)},g)
        $
        \item{} To assert $v$ is a linear combination of $\{v_i\}$, checks 
        $
        E(g_v^{\alpha_v v(s)},g) =
        E(g_v^{v(s)},g^{\alpha_v})
        $ the same is done for $w$ and $k$.
        \item{} To assert the same coefficients were used in each combination, checks 
        $
        E(g^Z,g^\gamma) =
        E(g_v{v^(s)},g_w{w^(s)},g_k{k^(s)},g^{\beta\gamma})$
      \end{itemize}
    \end{enumerate}
    The first two steps of the algorithm are computationally expensive but once the keys are calculated and published, any verifier can be convinced with few checks and without interacting with the prover anew. The correctness of this verification algorithm follows from cryptographical principles out of the scope of this paper, like \textit{homomorphic hiding} and the \textit{knowledge of coefficient} test. 
  }
}

\section{Future work} {
  Currently, the ZeroKnowledgeProofs paclet implements a compiler for a subset of Wolfram language into QAP, the generation of the evaluation, verification and witness keys is visualized as future work, along with the development of a different compiler for boolean circuits into a similar NP-complete problem called \textit{Quadratic Span Programs}.
}

%define the following sections to hide their Section Number (Notes Style)
\ledgernotes

%AUTHOR: comment out if using thebibliography
%\theendnotes

%AUTHOR: please read ledgerbib.bst usage notes by opening it in a text editor. We have modified it to include the use of the @misc item type for the proper formatting of online sources.

\bibliographystyle{ledgerbib}
\bibliography{ms}
\nocite{*}

%AUTHOR: comment out, this is used to make sure the Creative Commons License
%image fits on page

\newpage 	
%define the following sections to have the Appendix Style

\appendix
\setcounter{section}{0}

\section{ZeroKnowledgeProofs paclet usage}\label{apndxCode} {
  \vspace{-10pt}
  \includegraphics[width=0.87\textwidth]{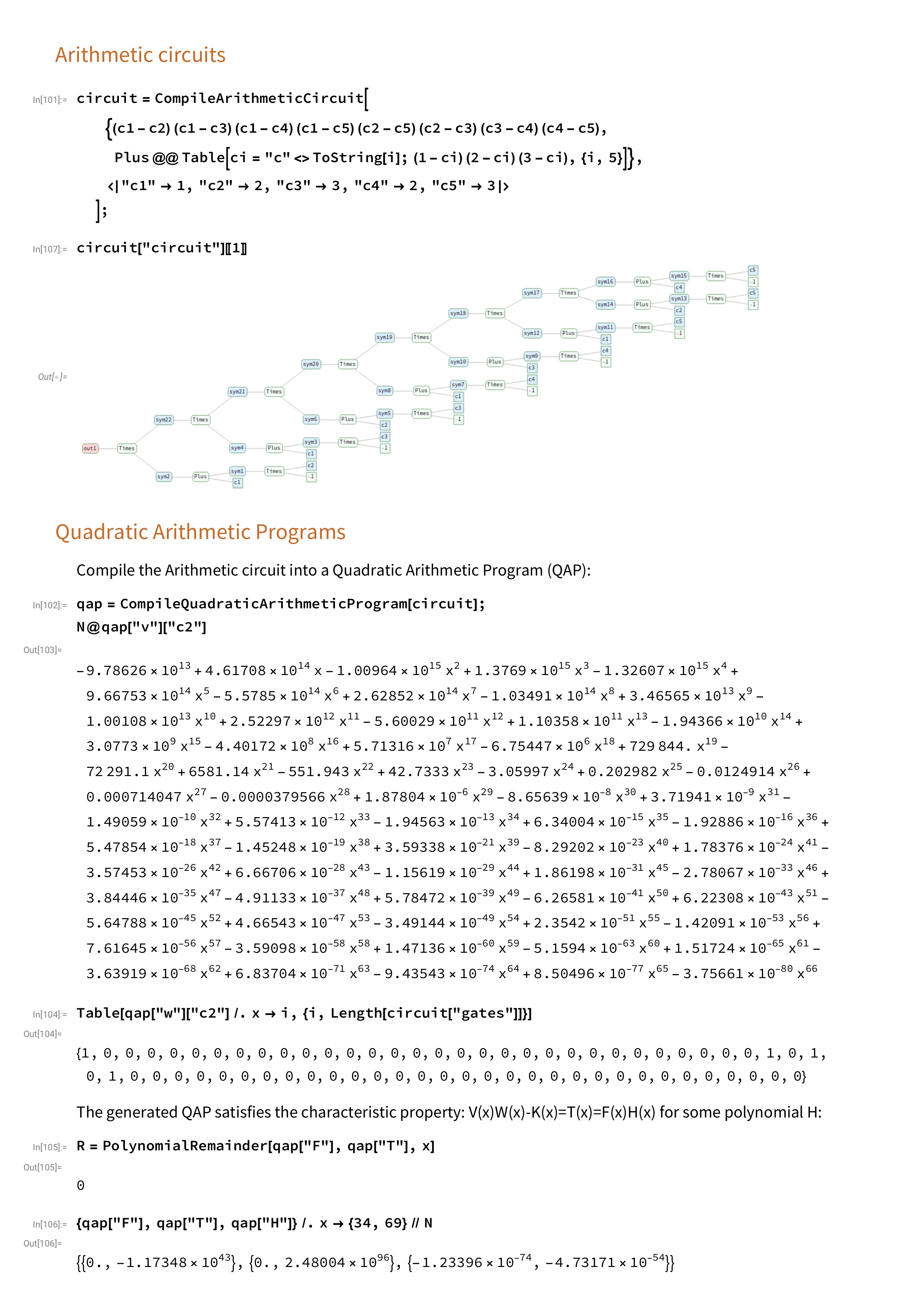}
} 
%\newpage
%here up^^

\thispagestyle{pagelast}

%\theendnotes

\end{document}